# Competing misfit relaxation mechanisms in epitaxial correlated oxides

Felip Sandiumenge, [1,*] José Santiso, [2] Lluis Balcells, [1] Zorica Konstantinovic, [1] Jaume Roqueta, [2] Alberto Pomar, [1] Juan Pedro Espinós, [3] and Benjamín Martínez[1]

[1]Institut de Ciència de Materials de Barcelona, CSIC

Campus de la Universitat Autònoma de Barcelona

Bellaterra, Catalonia 08193 (Spain)

[2]Centre for Nanoscience and Nanotechnology, CIN2 (CSIC-ICN)

Campus de la Universitat Autònoma de Barcelona

Bellaterra, Catalonia 08193 (Spain)

[3]Materials Science Institute of Seville, ICMS (CSIC-University of Seville)

c/Américo Vespucio 49, E-41092 Seville, (Spain)

*corresponding author: e-mail: felip@icmab.es

Abstract:

Strain engineering of functional properties in epitaxial thin films of strongly correlated oxides exhibiting octahedral-framework structures is hindered by the lack of adequate misfit relaxation models. Here we present unreported experimental evidences of a four-stage hierarchical development of octahedral-framework perturbations resulting from a progressive imbalance between electronic, elastic and octahedral tilting energies in La$_{0.7}$Sr$_{0.3}$MnO$_3$ epitaxial thin films grown on SrTiO$_3$ substrates. Electronic softening of the Mn - O bonds near the substrate leads to the formation of an interfacial layer clamped to the substrate with strongly degraded magnetotransport properties, i.e. the so-called dead layer, while rigid octahedral tilts become relevant at advanced growth stages without significant effects on charge transport and magnetic ordering.

Thin film epitaxy provides the common playground for tailoring materials functionalities through misfit strain. However, notwithstanding the huge attention focused on this issue [1,2], the lack of understanding of misfit relaxation mechanisms in octahedral framework structures still constitutes a serious drawback for a deterministic manipulation of the lattice, electronic and magnetic degrees of freedom which characterize them. Controlling octahedral tilts and distortions in perovskite-type transition-metal oxides has accordingly emerged as a critical step towards exploiting their unique capabilities for electronic and spintronic applications [3]. Notably, epitaxial perovskites often bypass misfit-dislocation mediated mechanisms that successfully describe the relaxation behavior of semiconducting epitaxial films [4]. Examples are found among a variety of functional films like ferroelectric PbTiO$_3$ [5], multiferroic BiFeO$_3$ [6] and TbMnO$_3$ [7], where misfit strains are relieved by a combination of symmetry changes and ferroelastic domains. Special interest is being attracted by the room temperature (T$_C$ ~ 370 K) half-metal ferromagnet

La$_{0.7}$Sr$_{0.3}$MnO$_3$ (LSMO) [3]. The intriguing degradation of its magnetotransport performance near the interface, constitutes, however, a serious drawback whose microscopic origin remains obscure. Prior reports converge in signaling a Mn$^{3+}$ enrichment near the interface [8,9] and a preferential occupation of *d* - $e_g$ $3z^2$ - $r^2$ orbitals [10-12], leading to a local *C*-type antiferromagnetic ordering. Disentangling the effects of the various constraints imposed by the interface (symmetry breaking [13], elastic strain [12], interaction between the Mn and Ti electronic structures [12] and polar discontinuity [14]) thus emerges as a cornerstone for the understanding of misfit relaxation mechanisms in correlated-oxide epitaxial framework structures.

To explain their elastic behavior, here we consider that the deformation of perovskite-type ABO$_3$ octahedral frameworks is governed by the relative strength between the B - O - B bond angles bridging adjacent BO$_6$ octahedra and the B - O bonds. Strain may couple with octahedral tilts [15,16], or with B - O bonds either at the expense of elastic or electronic energies. Strongly correlated oxides constitute an ideal arena to explore the effects of biaxial strain on the delicate balance between lattice, spin, charge and orbital degrees of freedom [17]. To address this issue, here we investigate the thickness dependence of lattice distortions in La$_{0.7}$Sr$_{0.3}$MnO$_3$ (LSMO) films grown on (001)-SrTiO$_3$ (STO) substrates.

Films with thicknesses ranging between 1.9 nm and 475 nm were grown by RF magnetron sputtering on TiO$_2$ terminated STO substrates [18]. Kinetic conditions were adjusted to guarantee a 2D layer-by-layer growth mode and avoid growth induced deterioration of the films [19]. The strain state was investigated by x-ray diffraction determinations of the out-of-plane (*c*) and in-plane (*a*) lattice parameters by fitting positions of both integer and half-order superstructure reflections in order to account for interferences between substrate and film reflections in the low thickness range (< 4 nm) [19]. The 3D shear strain state of the films was determined by measuring the in-plane ($\phi_\parallel$) and out-of-plane ($\phi_\perp$) components of the shear angle, $\phi_i = \alpha_{rh,i} - 90º$ [19]. Twin patterns were directly observed by orientation contrast electron backscattered (OC-EBS) images obtained in a field emission scanning electron microscope. The magnetic and charge transport properties of the films were studied between 10K and 380K. The in-plane electrical resistivity was measured using the standard four-point geometry with a constant applied current of 5 nA. Magnetization measurements were performed using a SQUID magnetometer under perpendicular applied field of 5 kOe. The Mn oxidation state was explored by means of X-ray photoemission spectroscopy (XPS).

The evolution of the lattice parameters with increasing film thickness is shown in Fig. 1(a). A progressive decrease of the *c*-axis parameter from the thinnest film ($h \sim 1.9$ nm) down to a minimum value at thicknesses comprised between 10 nm and 25 nm followed by a smooth increase is clearly seen. This result agrees with a dilation of the substrate-film interface, as reported by other authors [20]. Notably, this trajectory is accompanied by a constant in-plane fully strained *a*-axis parameter throughout the whole thickness range. In the full thickness range ½ ½ *L* crystal truncation rod exploration has shown half–order reflections at *L*=3/2 and 5/2 (absence at *L*=1/2) consistent with rhombohedral $a^-a^-a^-$ (*R-3c*), orthorhombic $a^-a^-c^0$ (*Imma*) and monoclinic $a^-a^-c^-$ (*C2/c*) tilt systems [21]. According to this evolution, we identify in Fig. 1(a) four different deformation regimes, namely: I) for $h < 2.5$ nm, II) for $2.5$ nm $< h < 10$ nm, III) for $10$ nm $> h > 25$, and IV) for $h > 25$ nm up to 475 nm. As depicted in Fig. 1(b), XPS of the Mn 2p3/2 peak shows a progressive shift of -0.2 eV towards lower binding energies as the film thickness is decreased below 10 nm, particularly within regime I. According to Abbate *et al.* [22], this shift would correspond to a variation of 0.18 in the oxidation state of Mn which would drive the LSMO phase into a non-ferromagnetic and insulating state [23]. The resulting evolution of the Poisson's ratio, $v = \varepsilon^\perp/(\varepsilon^\perp - 2\varepsilon_\parallel)$, where $\varepsilon^\perp$ and $\varepsilon_\parallel$ are the out-of-plane and in-plane strain components ($a = 3.881$ Å [24] is taken as a reference value), is shown in Fig. 1(c), and the evolution of the unit cell volume, *V(h)*, is depicted in Fig. 1(d). The shadowed area, where films exhibit auxetic behavior, indicates the non-twinned thickness range (see below). In regime III, $v \sim 0.33$, is the closest value to that derived from the elastic constants determined for a $La_{0.83}Sr_{0.17}MnO_3$ single crystal, $v = 0.41$[19, 25]. Concomitantly, the in-plane ($\phi^\parallel$) and out-of-plane ($\phi^\perp$) components of the rhombohedral shear, $\alpha_{rh} - 90°$, also converge in regime III [Fig. 1(e)] (recall that the rhombohedral structure imposes $\phi^{\parallel} = \phi^{\perp}$). Therefore, regime III corresponds to a pure elastically strained state of the rhombohedral LSMO film without invoking any significant octahedral tilting perturbation but a compression of the $MnO_6$ octahedra along the *c*-axis, in agreement with previous spectroscopic analyses [26]. In order to correlate the divergence between $\phi^{\parallel}$ and $\phi^{\perp}$ observed for $h > 10$ nm with the perturbation of the octahedral tilt pattern, here we use the formalism developed for rhombohedral perovskites [27] to derive an approximate analytical expression for the unit cell volume dependence on the in-plane, $\alpha$ (=$\beta$), and out-of-plane, $\gamma$, octahedral tilt angles about the [100]/[010] and [001] axes, respectively: $V \approx a_0^3 (1-\gamma^2)(1-\alpha^2)^2$, where $a_0$ is the equilibrium lattice parameter of the undistorted cubic prototype [19]. The resulting $V(h)$ dependence shows an excellent agreement with that calculated from the lattice parameters, as illustrated by a red solid line in Fig. 1(d). This indicates that the relaxation pathway followed in regime IV is mediated by a progressive decrease of $\alpha$ (=$\beta$), causing a simultaneous enlargement of the three principal axes of the unit cell, accompanied by a increase of $\gamma$ which has no effect on the *c*-axis parameter but compensates the effect on *a* (and *b*), thus preserving the in-plane matching with the substrate.

Turning now our attention to the very early growth stages, we illustrate in Fig. 2 the evolution of the shear strain state through film thicknesses of 1.9 nm (a) and 3.3 nm (b). The twin pattern is recognized in the OC-EBS image shown in Fig. 2(b), left panel, as a patchwork of twin families seen as parallel stripes of alternating dark/bright contrast aligned with the [100] and [010] film directions. Conversely, in the 1.9 nm film shown in Fig. 2(a), this contrast is absent. The speckled contrast is attributed to the coexistence of two degenerated monoclinic epitaxial orientations. The corresponding *HK* reciprocal space maps (RSM's) around the film and substrate 200 and film 1/2 1/2 3/2 reflections are shown in the right panels. For the 1.9 nm thick film, neither the 200 nor the 1/2 1/2 3/2 reflections exhibit any signature of twinning. In the thicker film, the diffuse scattering associated with the 200 reflection along the *K* axis defines up to second order modulation satellites signaling the development of the lateral twin periodicity [28]. Moreover, the sharp four-fold splitting along the [100]* and [010]* reciprocal directions observed around the 1/2 1/2 3/2 half-order reflection is consistent with the development of an homogeneous rhombohedral (100)/(010) twin structure as schematically depicted in Fig. 2(c). A similar analysis of a 2.7 nm thick film (not shown) revealed weaker twin signatures in the RSM's. These observations contrast with equilibrium models of ferroelastic domains in epitaxial LSMO predicting an exponential decay of the twin width with increasing thickness for $h$ < 3 nm [29], and indicate an spontaneous build up of the shear strain at a critical thickness, $h_{shear}$, comprised between 2 nm and 2.5 nm (5 - 7 unit cells).

The evolution of the temperature dependence of the resistivity, $\rho(T)$, at these growth stages is shown in Fig. 3(a). It can be seen that for $h$ < $h_{shear}$, the films are insulating without any signature of insulating-to-metal (I-M) transition between 10 K and 380 K. This behavior is accompanied by a sudden drop of the magnetic moment (see Fig. 3(b), main panel and lower inset). In concomitance with this behavior, the temperature dependence of the magnetization, $M(T)$, shown in the upper inset of Fig. 3(b) also shows that the films are not ferromagnetic (FM). According to our XPS analysis, see Fig. 1(b), this transition is driven by a local increase of the $Mn^{3+}/Mn^{4+}$ ratio near the interface, resulting in an enhanced Jahn-Teller distortion and a decrease of the band width. According to the structural $La_{1-x}Sr_xMnO_3$ phase diagram [30], this effect is analogous to a reduction of the doping level, x, which favors the stabilization of monoclinic forms of LSMO. In such a case, the in-plane shear strain is cancelled by placing the monoclinic unique axis out-of-plane. This interpretation agrees with the observation of a critical thickness for twin formation coinciding with the built-up of a shear strain. The value of $h_{shear}$ ~ 2.5 nm coincides with the thickness at which Mn cations start to recover their stoichiometric mean valence, as observed in Fig. 1(b). Moreover, besides the distortion induced by Jahn-Teller and possible $e_g$ ($3z^2 - r^2$) orbital reconstruction effects, the mean volume of the $MnO_6$ octahedra is expected to increase as a result of the larger $Mn^{3+}$ ionic radius (0.645 Å in high-spin state, $vs$ 0.530 Å for $Mn^{4+}$ [31]). This is clearly manifested by the average increase of the Mn - O bond distances in the monoclinic phase of LSMO [30]. This causes an increase of the lattice parameters [30], in excess to that required to fully accommodate the tensile misfit strain with the STO substrate. This scenario explains the observed delay in the formation of an elastically strained rhombohedral phase until the growing film reaches regime III.

Fig. 4 shows the thickness dependent phase diagram of LSMO films. The formation of a ~ 2.5 nm thick insulating interfacial dead layer (indicated by a shadowed area) lacking FM order is compatible with prior experimental [11] and theoretical [13] reports showing that irrespective of the elastic strain state a preferential occupation of $3z^2 - r^2$ orbitals is intrinsic to the symmetry breaking at LSMO film surfaces and interfaces. According to theoretical studies, above the shear transition the progressive decay of the $Mn^{3+}$ concentration would favor the development of inhomogeneous FM patches with variable $Mn^{3+}/Mn^{4+}$ ratios [32]. This scenario is consistent with the appearance of the I-M transition in the $\rho(T)$ curves for $h \geq h_{shear}$ [see Fig. 3(a)], although this occurs at a lower temperature than that corresponding to the bulk Curie temperature $T_C$. Regime II, bridging the monoclinic and rhombohedral states, thus defines a transition between the interfacial state and a strained state in which the electronic energy gained by removing the orbital degeneracy within the $Mn^{3+}O_6$ coordination environment becomes progressively exceeded by the elastic energy opposing a similar expansion of the equatorial Mn - O distances, until the tensilely strained rhombohedral phase condenses at a thickness of ~ 10 nm (onset of regime III). Simultaneously, the typical temperature dependence of the magnetization in a FM material becomes apparent in the $M(T)$ curves shown in the upper inset of Fig. 3(b). Upon increasing the film thickness beyond regime III (~ 25 nm), the stored elastic energy is relaxed by a pure octahedral tilting mechanism without detrimental consequences on the metallic and ferromagnetic behavior of the films. The smooth decrease of $T_C$ observed at this stage [see Fig. 3(c)] can be attributed to a weakening of the magnetic interactions caused by the tilting perturbation.

In summary, the present work sheds light on the precise role of octahedral tilts and distortions in the misfit-strain relaxation behavior of an epitaxial octahedral-framework correlated-oxide structure. Our results bring up a new misfit relaxation scenario in thin film epitaxy whereby plastic deformation is bypassed by the succesive stabilization of characteristic deformation states that satisfactorily explain the intriguing thickness dependence of functional properties observed in LSMO films, including the formation of the so-called interfacial dead layer.

We thank J. Bassas (SCT, Universitat de Barcelona) and P. Ferrer (European Synchrotron Radiation Facility, ESRF-Spline, France) for their valuable support in X-ray diffraction experiments. We acknowledge the Spanish Ministerio de Economia y Competitividad and Consejo Superior de Investigaciones Científicas for financial support and for provision of synchrotron radiation facilities in using beamline BM25-SpLine". This research was supported by Spanish MEC (MAT2009-08024 and MAT2011-29081-C02), CONSOLIDER (CSD2007-00041 and CSD2008-00023), and FEDER program. Z. K. thanks the Spanish MEC for the financial support through the RyC program.

Supporting Information is available online from Wiley InterScience or from the author.

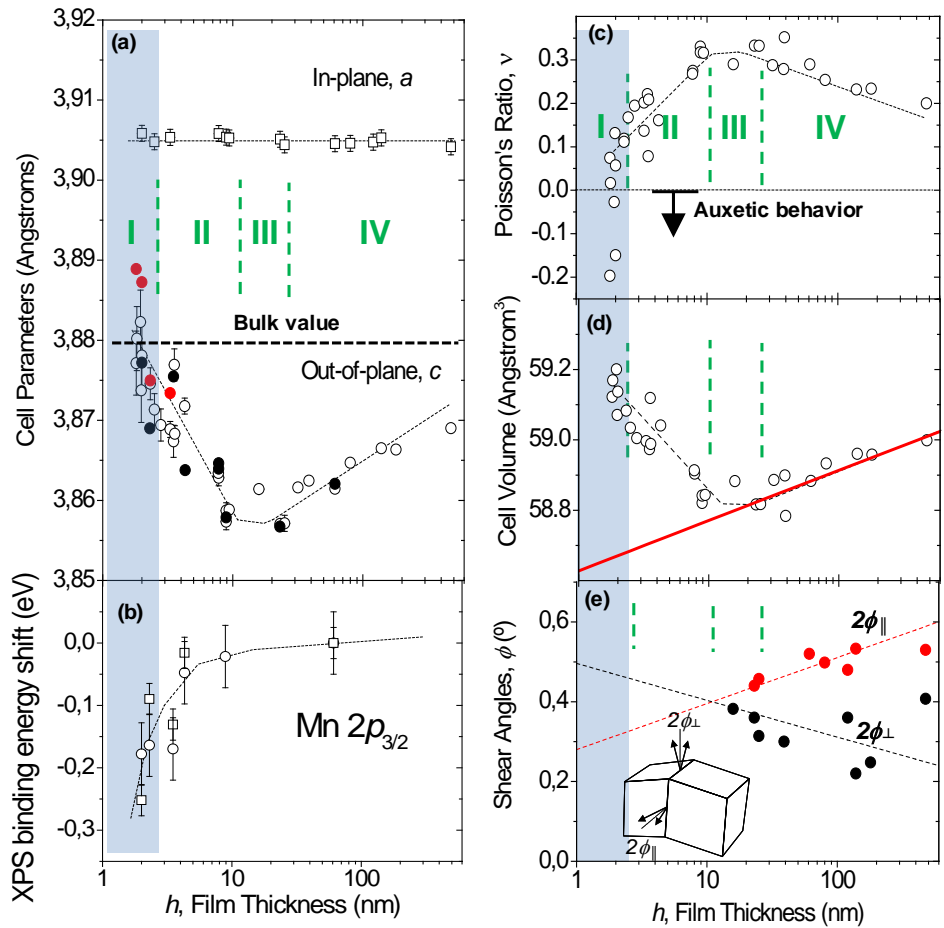

**Figure 1.** (a) Thickness dependence of the in-plane and out-of-plane lattice parameters. For *c*-axis parameters, open symbols were obtained by fitting 002 rocking profiles, black filled symbols from 1/2 1/2 3/2 superstructure reflections, and red symbols by averaging over 17 different $h/2\ k/2\ l/2$ ($h,k,l$ =1, 3, 5) superstructure reflections [19]. (b) Shift in the XPS binding energies of the Mn $2p_{3/2}$ peak for LSMO films with different thicknesses, relative to the 61nm thick sample. Circles and squares correspond to measurements performed with Escalab and Phoibos equipments, respectively [19]. (c) Thickness dependence of the Poisson's ratio. (d) Thickness dependence of the *pseudo*cubic unit cell volume. (e) Thickness dependence of the in-plane ($\phi_\parallel$) and out-of-plane ($\phi_\perp$) shear angles. The shadowed area indicates the non-twinned thickness range.

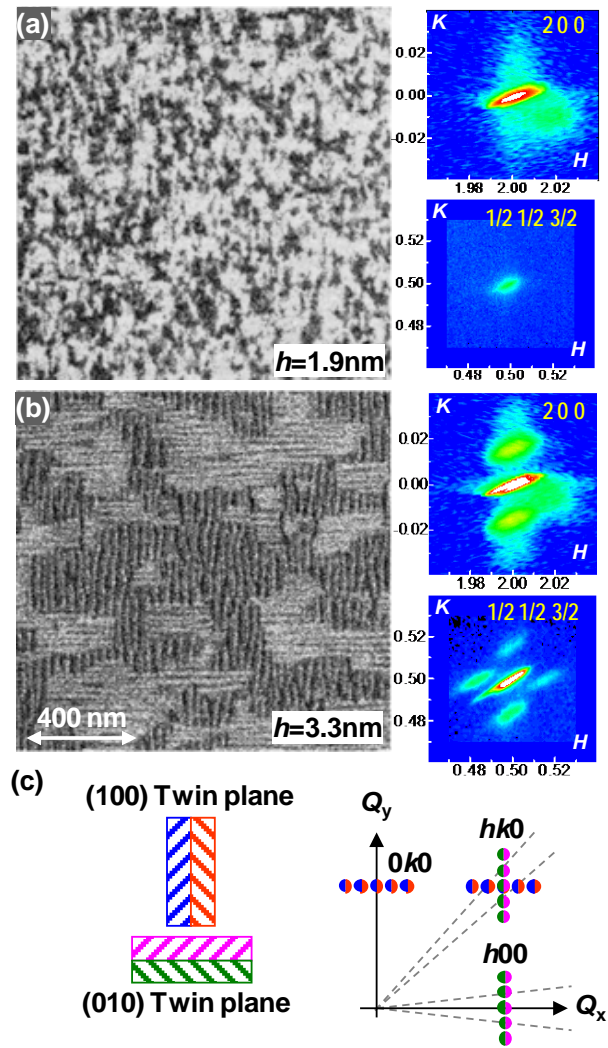

**Figure 2.** OC-EBS images corresponding to 1.9 nm (a) and 3.3 nm (b) thick films. Right panels show the corresponding *HK* RSM's of the 200 and 1/2 1/2 3/2 reflections. (c) Schematics showing the effect of (100) and (010) twin planes on diffraction.

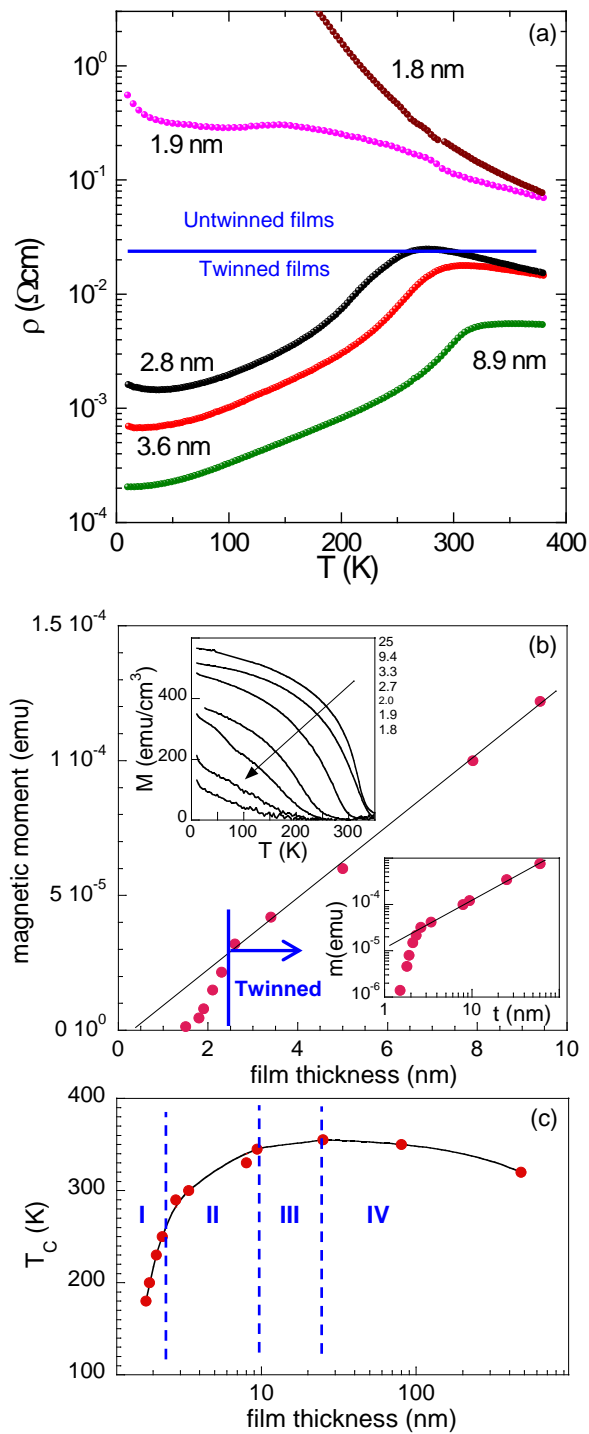

**Figure 3.** (a) Resistivity *versus* temperature curves for films having different thickness above and below the shear transition thickness, as indicated. (b) Thickness dependence of the magnetic moment for *h* < 10 nm. Lower inset shows the dependence in an expanded thickness range. Upper inset: Temperature dependence of the magnetization at low field (H = 5kOe). (c) Thickness dependence of the Curie temperature, $T_C$.

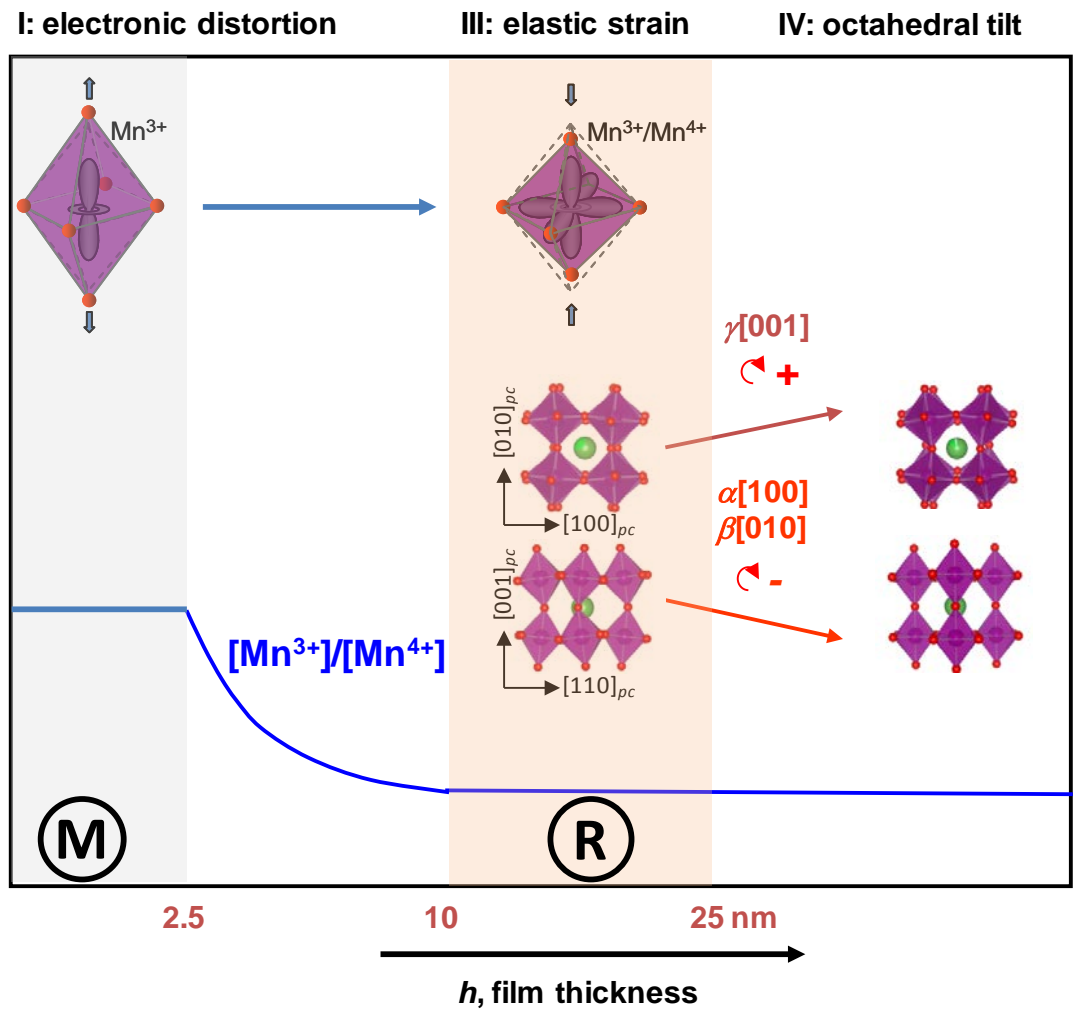

**Figure 4.** Mechanisms of lattice distortions through regimes I,II, III and IV. Regime II bridging the monoclinic and rhombohedral states, is not labelled in the diagram for clarity. The orbital character and its relation with octahedral strain in the monoclinic (M) and rhombohedral (R) states is indicated. The transition between the two states is mediated by a decrease of the $Mn^{3+}/Mn^{4+}$ ratio (blue line). Lower panels illustrate the combined octahedral tilting mechanism releasing the elastic energy stored in regime III.

SUPPLEMENTAL MATERIAL

# Competing misfit relaxation mechanisms in epitaxial correlated oxides

Felip Sandiumenge, [1,*] José Santiso, [2] Lluis Balcells, [1] Zorica Konstantinovic, [1] Jaume Roqueta, [2] Alberto Pomar, [1] Juan Pedro Espinós, [3] and Benjamín Martínez[1]

[1]Institut de Ciència de Materials de Barcelona, CSIC

Campus de la Universitat Autònoma de Barcelona

Bellaterra, Catalonia 08193 ( Spain)

[2]Centre for Nanoscience and Nanotechnology, CIN2 (CSIC-ICN)

Campus de la Universitat Autònoma de Barcelona

Bellaterra, Catalonia 08193 ( Spain)

[3]Materials Science Institute of Seville, ICMS (CSIC-University of Seville)

c/Américo Vespucio 49, E-41092 Seville, (Spain)

*corresponding author: e-mail: felip@icmab.es

## A. Growth of films

La$_{0.7}$Sr$_{0.3}$MnO$_3$ (LSMO) films with thicknesses ranging between 1.8 nm and 475 nm were grown on (001)-SrTiO$_3$ (STO) by RF magnetron sputtering from a LSMO stoichiometric ceramic target. Substrates were cleaned in an ultrasonic bath with milli-Q water and annealed at 1000 °C in air for 2h in order to obtain clean and smooth surfaces mostly terminated by TiO$_2$ atomic planes. After annealing, the substrate morphology exhibited the typical morphology of terraces and steps with heights corresponding to the STO unit cell ($a$~0.39 nm). The films were then grown at 900 °C in an oxygen pressure p = 0.19 mbar and subsequently in situ annealed at 900 °C under an oxygen pressure of 0.47 bar for 1 h to improve their structural, magnetic and transport properties [1]. These growth conditions ensured a layer-by-layer growth without any island formation, as inferred by inspection of atomic force microscopy images. This is an essential condition to avoid growth induced degradation of the structural, transport and magnetic properties of the films [2]. Films exhibited bulk-like saturation magnetization values of about 585 emu/cm$^3$ (above a thickness of 2.5 nm, as shown in Fig. 3b, in the main text), which is an indication of their high quality. X-ray reflectometry was used to determine their thickness.

## B. XRD procedures and determination of lattice parameters

X-ray diffraction measurements were carried out using three different experimental setups: (1) An in-house four-angle diffractometer equipped with a Cu-K$\alpha$ radiation source supplying X-rays with $\lambda$ = 1.5406 Å (X'Pert Pro MRD - Panalytical). We used a high resolution primary optics consisting of a parabolic mirror and a 4 x Ge(220) crystal asymmetric monochromator; (2) The same in-house diffractometer equipped with an X-ray lens optics at the primary beam and a Co-K$_\alpha$ radiation source ($\lambda$ = 1.7903 Å), in order to increase intensity of weak superstructure reflections of the type H/2 K/2 L/2 reflections [3]. (3) For the thinnest films (1.8 - 3.5 nm) a set of about 17 $\frac{H}{2} \frac{K}{2} \frac{L}{2}$ (*H, K, L*= 1, 3, 5) superstructure reflections were also measured by using a synchrotron radiation source (6-circle diffractometer BM25 - SpLine at the European Synchrotron Radiation Facility (ESRF), with an energy E = 15 KeV ($\lambda$ = 0.851 Å) at a fixed incidence angle of 0.5º slightly above the critical angle of 0.20º in order to maximize the intensity). Fig. S1 shows the 1/2 1/2 *L* crystal truncation rod superstructure reflections corresponding to a 3.5 nm thick film. Similar results were obtained in the whole thickness range. The appearance of the *L* = 3/2 and 5/2 reflections, along with the absence of *L*=1/2 is consistent with rhombohedral $a^-a^-a^-$ (*R-3c*), orthorhombic $a^-a^-c^0$ (*Imma*) and monoclinic $a^-a^-c^-$ (*C2/c*) tilt systems [4].

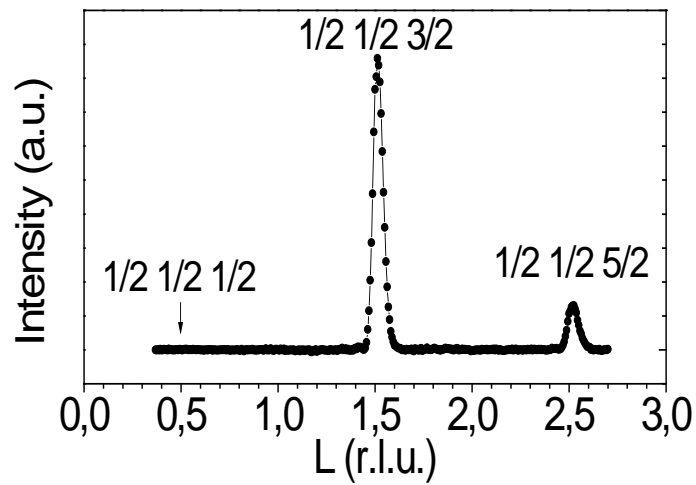

**Fig. S1.** *L*-scan along ½ ½ *L* rod of a 3.5 nm thick LSMO film indicating the presence of L= 3/2 and 5/2 half-order reflections, but the absence of L=1/2.

For the determination of the out-of-plane *c*-axis lattice parameter three strategies were adopted. First, they were extracted by fitting the 002 STO+ 002 LSMO profiles (open symbols in Fig 1a) (see Fig. S2). Since convolution with the intense substrate 002 reflection appeared critical to the precise determination of the 002 film peak position, especially for the thinnest films [5], we also analyzed the angular positions of non-convoluted $\frac{H}{2} \frac{K}{2} \frac{L}{2}$ superstructure asymmetric reflections arising from the distorted LSMO structure, thus allowing a direct determination of their angular positions. The interplanar distance of the most intense $\frac{1}{2}\frac{1}{2}\frac{3}{2}$ reflection was determined and the *c*-axis parameter was extracted assuming total in-plane matching with the STO substrate, as proven valid for the full set of analysed thicknesses (full black symbols in Fig. 1a). Additionally, for the thinnest films (1.8 nm - 3.5 nm) a set of about 17 $\frac{H}{2}\frac{K}{2}\frac{L}{2}$ (*H, K, L*= 1, 3, 5) superstructure reflections were also measured by using a synchrotron radiation source. The average *c*-axis parameter is depicted by red symbols in Fig 1a. We obtained consistent results by all three procedures.

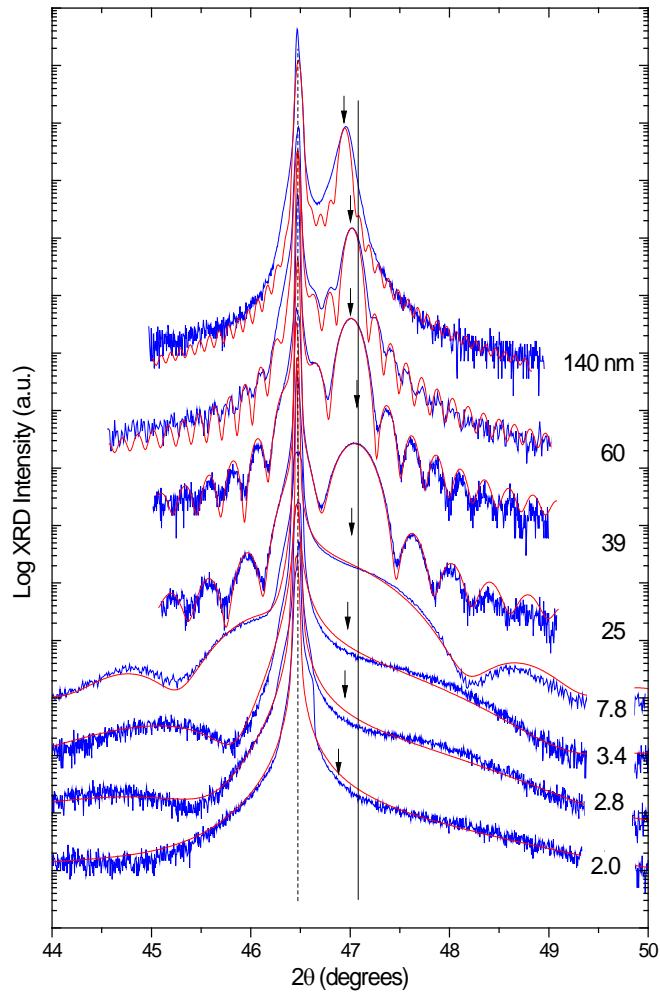

**Fig. S2.** High resolution rocking scans about the 002 reflections of LSMO film and STO substrate for films having different thicknesses, as indicated. Kiessig fringes correspond to the film thickness. The arrows indicate the position of the film reflections obtained from fitting the full curve profiles (red solid lines).

In-plane *a*-axis lattice parameters were determined either from reciprocal space maps around the -303 LSMO reflection for thick films (*h* >10nm), as the one depicted in Fig. S3 (left panel), or from in-plane $\phi - 2\theta$ diffraction maps of the 200 LSMO reflection for the thinnest films.

The right panel of Fig. S3 shows a schematic representation of the effect of (100) twinning on the reciprocal space around the 303 reflections for the rhombohedral *R-3c* structure. The 303 reflections are split only along *Qx*, whereas the 033 reflections are split along *Qx* and shifted to positive or negative *Qz* values. The empty symbols result from an in-plane 180º rotation about *Qz*. Assuming that epitaxial growth on the SrTiO$_3$ (001) cubic substrate generates evenly distributed *a,b* twin domains with 90º rotations about the film normal (i.e. preserving the average four-fold symmetry of the substrate), the 303/033 reflections will split around the *pseudo*cubic nodes into six contributions. Therefore, in the experimental *Qx,Qz* map shown in Fig. S3 (left panel) each of the two Bragg reflections having the same $Q_x$ value and split along the vertical direction in $Q_z$ correspond, in fact, to a double peak also split in $Q_y$ leading to six equivalent reflections. The in-plane lattice parameters may therefore be extracted from the common $Q_x$ coordinates and were found to be perfectly clamped to the substrate in the whole thickness range, even for this 475 nm thick film.

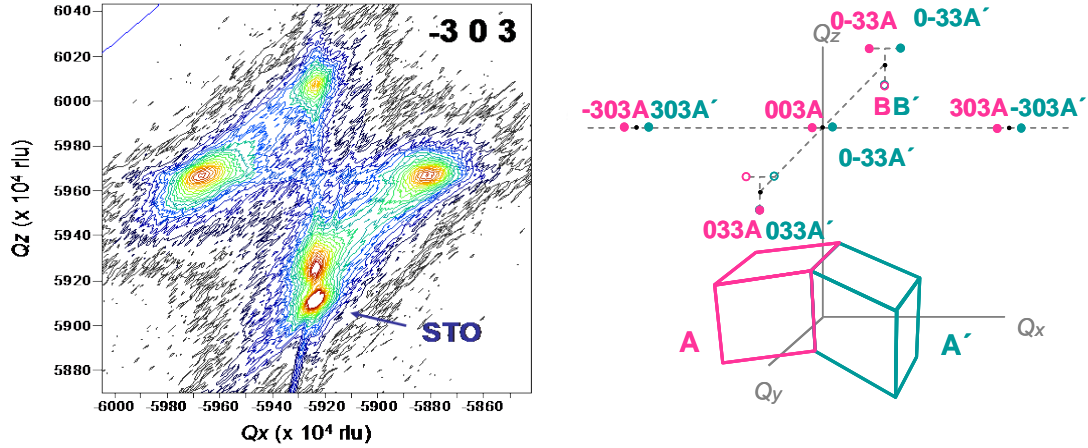

**Fig. S3**. Reciprocal space map of the -303 reflections (left panel) for the thickest LSMO film of 475 nm. The LSMO presents four different contributions forming a diamond shape. Their central position corresponds to the average in-plane parameter, fully coincident with substrate $Q_x$ position. The right panel shows a reconstruction of reciprocal space including reflections for *A* and *A'* [ and *B/B'*] twin rhombohedral domains with common (100) [ or (010)] twin plane.

**C. Determination of the shear strain state of the films**

The shear distortion of the rhombohedral LSMO structure, referred to its cubic aristotype, is given by the rhombohedral angle, $\alpha_{rh}$, which is related to the twin angle by $\phi = 2(\alpha_{rh} - 90°)$. The in-plane ($\phi^{\parallel}$) and out-of-plane ($\phi^{\perp}$) components of the twin angle, were determined by combining X-ray diffraction measurements over out-of-plane and in-plane split components of well-developed 002 and 200 Bragg reflections, respectively. Note that in the rhombohedral structure ($\phi^{\parallel} = \phi^{\perp}$). Therefore, any divergence between these two values indicates an anosotropic shear strain state of the film and a deviation from the rhombohedral distortion. In-plane $2\theta/\omega$-$\phi$ area scans of 200 reflections were measured with the in-house XRD equipment using a non-monochromatic Cu-$K\alpha$ primary beam with grazing incidence and exit angles of about 0.5º, and a parallel plate collimator of 0.27º acceptance angle for the in-plane diffracted beam. For the out-of-plane 002 reflection we performed $Q_x$ horizontal scans along a direction parallel to the [100] in-plane substrate axis, at $Q_z$ values corresponding to the 00$L$ ($L$=1,2,3) LSMO reflections. For these measurements we used a high resolution optics and Cu-$K\alpha_1$ radiation supplied by a 4xGe(220) asymmetric monochromator. In some cases the lateral (in-plane) correlation of the twin structure was also extracted from the same measurements. Typical examples of both types of measurements are depicted in Fig. S4 showing clear differences between thin ($h$ < 10nm) and thick films ($h$ <10nm). In-plane diffraction area scans of the 200 reflection for thick films reveal a clear splitting coming from Bragg peaks arising from the in-plane component of the twin angle. In the map depicted in Fig. S4 (left, top) the twin angle $\phi^{\parallel}$ corresponding to a 140 nm film thickness was about 0.53º. For thin films thinner than 10 nm the in-plane area scans show satellite peaks coming, not from domain tilting, but from the twin periodicity. In the case depicted in Fig. S4 (left, bottom) the 9 nm thick film shows up to 2nd order satellites corresponding to a twin periodicity of Λ=39 nm (consistent with the SEM observations for this sample). Similarly, the out-of-plane 00$L$ reflections may reveal features coming either from the Bragg contributions of tilted twin domains or from twin periodicity. For a given 00$L$ reflection the weight of either one or the other contribution depends on a number of factors, such as film thickness, twin lateral periodicity Λ and tilt angle $\phi$, along with $2\theta$ angle [6]. As it is observed in the $Q_x$ scans of the thick films (140 nm) the 001 reflection shows clear satellite peaks up to 2nd order, which correspond to a twin lateral periodicity of $\Lambda$= 146 nm, consistent with the SEM observation of the same sample. On the other hand, the 002 and 003 scans show broader lateral peaks corresponding to Bragg reflections of tilted twin domains (with tilt angle of $\phi\perp$ = 0.23º, remarkably different to the corresponding $\phi\parallel$ value of 0.53º). Unfortunately, for films thinner than 10 nm we could not resolve the two Bragg contributions arising from the misorientation between domains along $Q_x$ and $Q_y$ from the satellite peaks produced by the lateral modulation of the twin planes, precluding a precise determination of the twin angles.

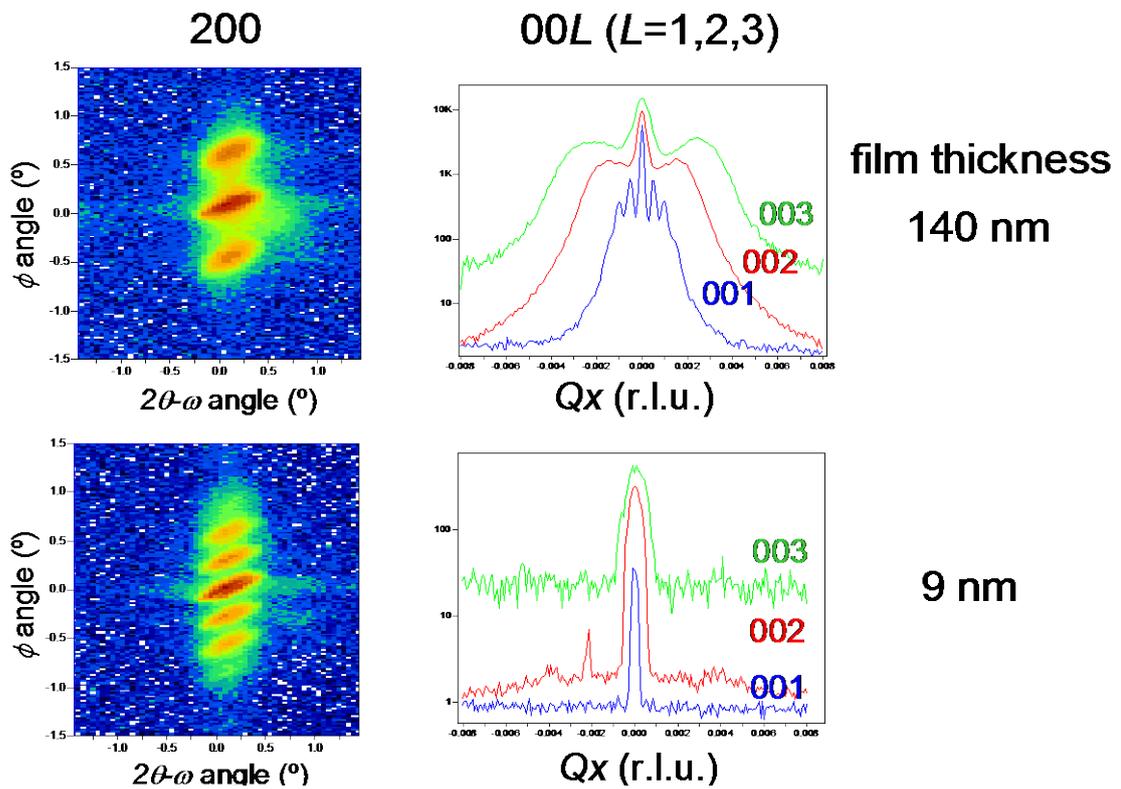

**Fig. S4.** Left panel shows the in-plane $\phi$-$2\theta$ area scans around the 200 LSMO reflection for a thick LSMO film of 140 nm (top) and a thin film of 9 nm (bottom). Right panel shows the corresponding out-of-plane reciprocal space horizontal ($Qx$) scan of 00$L$ LSMO reflections ($L$=1,2,3). For the thick film (top) $L$=2 and 3 reveal the Bragg peaks arising from twin domains related by (100) twin planes, whereas for $L$=1 satellites are related to twin domain periodicity. No significant satellite signal was observed for the $Qx$ scans of the thinnest film (bottom).

**D. Calculation of unit cell volume variations from a pure octahedral tilt model**

The *calculation of cell volume variations from a pure octahedral tilt model,* depicted in Fig. 1(d), is based on the experimentally determined dependence of $\phi^{\parallel}$ and $\phi^{\perp}$ on film thickness [Fig. 1(e)]. Assuming a rigid-octahedron model, in the rhombohedral structure any slight deviation from the equilibrium octahedral tilt pattern is directly correlated with changes in the twin angles. Megaw and Darlington derived a simple analytical expression relating the rhombohedral angle (referred to the *pseudo*cubic cell), $\alpha_{rh}$, and the octahedron tilt angle, $\omega$ [7],

$$\cos \alpha_{rh} \cong \frac{\sin^2 \omega}{3 - 2\sin^2 \omega} \quad (1).$$

Given that for the equilibrium *R-3c* La$_{0.7}$Sr$_{0.3}$MnO$_3$ structure $\omega \sim 7°$ and $(90° - \alpha_{rh}) \sim 0.4°$, for a qualitative calculation we may assume $\sin\omega \sim \omega$, so equation (1) becomes $(90° - \alpha_{rh}) \approx (1/3) \omega^2$. If changes in tilt angles are small enough we may establish separate in-plane and out-of-plane tilt components following

$$\phi_{//} \approx \tfrac{1}{3}\gamma^2$$

$$\phi_{\perp} \approx \tfrac{1}{3}\alpha^2 \quad (2)$$

respectively, being $\alpha$ ($=\beta$) and $\gamma$, the octahedron tilt angles about the [100], [010] and [001] directions, respectively. If we assume a rigid-octahedron model, the *pseudo*cubic in-plane, $a$ ($=b$), and out-of-plane, $c$, lattice parameters depend on the octahedral tilt components as:

$$c = a_0 \cos^2 \alpha$$

$$a = a_0 \cos \alpha \cos \gamma \quad (3).$$

Therefore the unit cell volume can be expressed as,

$$V = a^2 c = a_0^3 (\cos^2 \gamma)(\cos^4 \alpha) \approx a_0^3 (1-\gamma^2)(1-\alpha^2)^2 \quad (4)$$

where $a_0 = 2\, d_{Mn-O} = 3.935$ Å is the size of the regular octahedron. Using the $\gamma$ and $\alpha$ values calculated from the linear dependence of the measured $\phi^{\parallel}$ and $\phi^{\perp}$ angles depicted in Fig. 1(e) in equation (4), and assuming the reference unit cell volume as the one measured for the 10-25 nm thick films ($V_{ref} = 58.8$ Å³), the unit cell volume variation depicted as a continuous line in Fig. 1(d) was obtained. The result is in perfect agreement with experimental values for thicknesses $h > 10$ nm, while it clearly diverges for $h < 10$ nm if an extrapolation of the same behaviour of $\phi^{\parallel}$ and $\phi^{\perp}$ was assumed for this region. This indicates that in region III the evolution of the unit cell distortion is fully consistent with a pure octahedral tilting behaviour.

**E. Comparison between thin-film and single-crystal values of the Poisson's ratio**

Poisson's ratios derived from strain values in thin films (present work and Ref. 8 are typically lower than those determined from single crystal measurements [9]. This discrepancy can in principle be attributed to the difference in La/Sr ratio (and therefore the $Mn^{3+}/Mn^{4+}$ ratio) between the single crystal used for the determination of the elastic constants and our films. Darling *et al.* measured $C_{11} = 225.5$ GPa, $C_{12} = 157.4$ GPa for the rhombohedral phase of $La_{0.83}Sr_{0.17}MnO_3$ [9]. According to those authors, the cubic symmetry ($\nu = C_{12}/(C_{11}+C_{12})$) yields the better description of the elastic response of this phase [9]. The good fit to cubic symmetry can be attributed to the occurrence of twins along all crystallographic planes allowed by the $m3m \rightarrow -3m$ ferroelastic transition, namely {100} and {110}, which produce an apparent macroscopic cubic symmetry (see e.g. Ref. 10). On the other hand, the in-plane shear induced by epitaxy in the present films results in a selection of only the {100} twin family perpendicular to the substrate. We believe that this is the main source of discrepancy when comparing the elastic behavior of thin films and single crystals.

**F. Determination of Mn oxidation state variations by X-ray Photoemission Spectroscopy**

The XPS photoemission spectra of core Mn 2$p$ level were measured for the same series of LSMO/STO films with different thicknesses. We used two different equipments: an Escalab 210 (Vacuum Generators) and a Phoibos 150 (Specs), which are equipped with monochromated and unmonochromated Al K$\alpha$ radiations, respectively. The pass energy values used in these instruments were 50 eV (Escalab 210) and 20 eV (Phoibos 150), and the energy resolutions, as measured by the FWHM of the Ag $3d_{5/2}$ peak for a sputtered silver foil, were 1.55 eV and 1.0 eV, respectively. In all the cases, as the samples are quite insulating, the binding energy scale was calibrated with the La $3d_{5/2}$ peak at 833.70 eV. The Shirley algorithm [11] was applied for the background subtraction. Determination of the Mn oxidation state from the Mn 2$p$ photoemission signal is particularly difficult because of the similarity in shape for $Mn^{+4}$ and $Mn^{3+}$ species. However, it is still possible to infer some qualitative trends when comparing spectra from samples obtained under the same conditions. The Mn 2$p$ spectra for a series of LSMO/STO films with different thicknesses, obtained with monochromated Al $K\alpha$ are plotted in Fig. S4 (a) after normalizing their intensities. Quite similar spectra, but a bit wider, were obtained with the unmonochromated Al $K\alpha$ source (not shown).

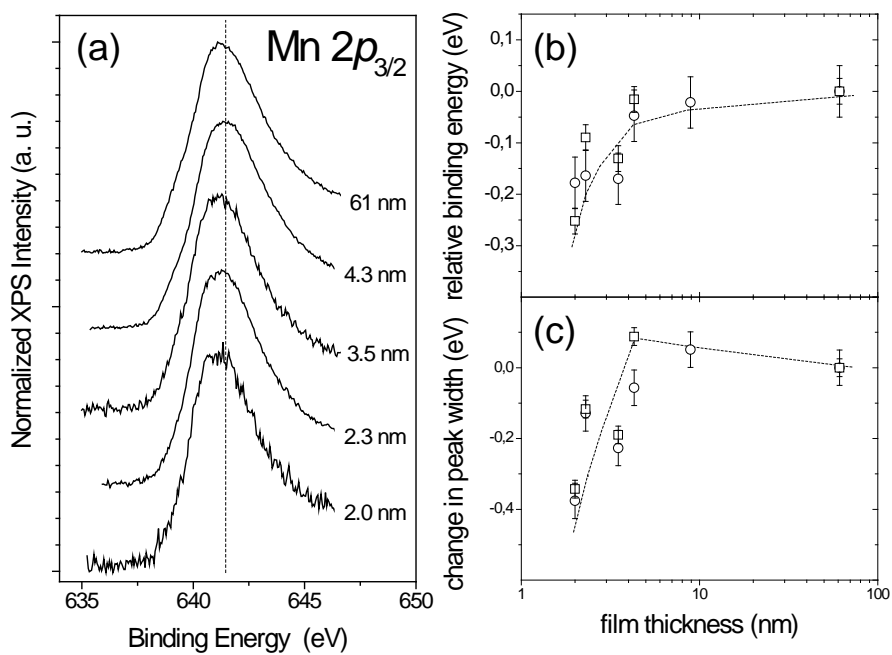

**Fig. S5**. (a) Detail of the Mn 2$p_{3/2}$ XPS peak for LSMO films with different thickness obtained with monochromated Al *K*α radiation. (b) Relative variations of the centroid position of the Mn 2$p_{3/2}$ peak distribution, and (c) corresponding full width half maximum changes. (The symbols correspond to the measurements performed with the two different equipments: Escalab: circles; and Phoibos: squares).

After background correction, the centroid of the lower binding energy Mn 2 $p_{3/2}$ contribution, along with the corresponding full width half maximum, were extracted from all the spectra in the range between 635 and 647.5 eV. The binding energy and FWHM of the Mn 2$p_{3/2}$ peak obtained for the thicker sample were taken as reference values and only relative variations were depicted. Figure S5 (*b,c*) shows both the relative binding energy shift as a function of film thickness as well as the FWHM variations. Although the position of Mn 2$p_{3/2}$ peak maxima do not show a substantial variation, their centroid positions clearly show a progressive shift to lower binding energies upon reducing film thickness, particularly below 10 nm, while at the same time the peak becomes narrower. The difference in binding energies between the 61 nm and the 2 nm films is about -0.2 eV. This is consistent with the shift to lower binding energies for the Mn $2p_{3/2}$ peak observed in $La_{1-x}Sr_xMnO_3$ bulk samples [12] when the $Mn^{3+/4+}$ mixed oxidation state is reduced by lowering the Sr content, $x$. In that work the Mn $2p_{3/2}$ peak shifts from 642.8 eV in $La_{0.1}Sr_{0.9}MnO_3$ to 641.8 eV for $LaMnO_3$, nominally corresponding to ratios $Mn^{3+}/Mn^{3+/4+}$ = 0.1 and 1.0, respectively. Assuming, as a first approximation, a linear dependence between the $Mn^{3+}$ ratio and the peak shift, a shift of 0.2 eV would roughly correspond to an increase of the $Mn^{3+}$ ratio of 0.18 (equivalent to a decrease of 0.18 in the Sr doping ratio). This shift would cause a progressive weakening of the ferromagnetic order and metallic character, as observed in the thinner films.

We thank J. Bassas (SCT, Universitat de Barcelona) and P. Ferrer (European Synchrotron Radiation Facility, ESRF-Spline, France) for their valuable support in X-ray diffraction experiments. We acknowledge the Spanish Ministerio de Economia y Competitividad and Consejo Superior de Investigaciones Científicas for financial support and for provision of synchrotron radiation facilities in using beamline BM25-SpLine". This research was supported by
Spanish Spanish MEC (MAT2009‐08024 and MAT2011-29081-C02-02), CONSOLIDER-INGENIO (CSD 2007−00041 and CSD 2008-00023), and FEDER program. Z. K. thanks the Spanish MEC for the financial support through the RyC program.